# Improving SDN Performance Using Network Coding: A Quantitative Analysis

Amer T. Ali, *Electronic Engineering Department, College of Electronics Engineering, Ninevah University*
*Email: amer.ali@uoninevah.edu.iq*
Qutaiba I. Ali, *Department of Computer Engineering, College of Engineering, University of Mosul*
*Email: Qut1974@gmail.com*

*Abstract—* **Software Defined Networking or SDN is an architectural approach to managing the network where the control and forwarding are different planes that are controlled through an application interface. Nevertheless, SDN's traditional renditions incorporate a few limitations such as packet drops, higher latency and suboptimal resource consumption especially in conditions of high volatility. These problems can reduce the benefits of SDN especially in areas that need large bandwidth and dependable networks. Network coding is a promising technology to encode and transmit the data packets as combinations and thus offers unique solutions to these problems. In communications, network coding, supports direct and efficient transmission and interactivities, smart data retrieval through the SDN-based networks enhancing both throughput, decreased latency, and effective fault tolerance. Further, it supports fine-tuning of the load distribution process by responding to network traffic situations and improving general networking capacity. In this paper, we investigate the applicability of applying network coding in SDN solutions and evaluate their effects. The evaluating shows that the network coding implemented reduces the value of packet loss and increases the value of delivery of data. This approach, allows for a 42.8% throughputs improvement boost under high load and yet does not incur high latency and is very fault tolerant. These results prove the possibility of network coding as a powerful addition for SDN and indicate its applicability in the future network structures.**

## I. INTRODUCTION

Software-Defined Networking (SDN) has changed the approach to networks design, management and operations. The control plane defines how information is directed within a network, while the data plane is responsible for routing the information; SDN makes the former centralized and more easily managed. This change of view shifts some network control and reconfiguration into software systems and places enhanced flexibility in the hands of network administrators. Considering the level of programmability that is inherent with SDN, it is easier to introduce new services, manage the traffic and to provide simplified network architecture. In addition, centralized structure of SDN provides end-to-end vision of the network that helps to optimize resources, enforce policies with higher accuracy and better protection of the network.[1][3]

Nevertheless, SDN has threats that may affect the performance in a dynamic nature and high bandwidth model. A primary challenge is packet loss that may result from congestion, link failure or a suboptimal routing decision. Packet loss distorts the reliability of data transmission while also requiring resend, adding pressure to network resources. Network congestion is another major problem, which appears in cases when traffic demand outperforms network capacity with decreased latency rates and less throughput. The classical solutions SDN use where the routing policies are previously set or configured in advance do not allow reacting to changeable network conditions and increase the negative consequences of congestion. Finally, availability is a significant issue due to scaling, which creates many prospects for single points of failure implicit within the SDN architecture.[1][4]

The management of challenges found in networks calls for strategies that taps into utilization of Software Defined networking and its control over networks in the most efficient way possible. One of such possible solution is known as network coding which potentially improves SDN's reliability and performance.[1]

Network coding combines several data packets into a single transmission with tuples which can be delivered instantaneously and merely via the intermediate nodes. It optimizes the throughput rate and enhances fault tolerance as this method does not slow down the processes as many people would presume. Through coding at appropriate places in the network, network coding allows users to recover lost packets from other parts of the network instead of having to request retransmission from the source. This lessens the amount of bandwidth used and also minimizes the extent of packet Drop. Further, through network coding, load can be shifted flexibly through traffic flows and is therefore helpful in preventing congestion and enhancing the network.[2][5]

Apart from these reliability advantages, network coding also has great advantages in terms of resource usage and expansibility. Through coded packets it means it simplifies the network bandwidth requirement in such a way that it transmits information about the original packets in a single transmission, unlike the normal packets that frequently transmit data to all the intended recipients. This is especially advantageous when the data is going to be received by many users using the multicast mode. Further, the integration of network coding into SDN increases programmability of the SDN architecture in such a manner that, adaptation of coding strategies according to the conditions on the network, the traffic intensity, and present applications is done in real-time.[2][6]

Although network coding gives much potentiality, the integration of network coding in SDN scenario raises some critical issues about challenges and drawbacks. The procedures for encoding and decoding also add certain costs, which can become noticed in terms of network response time and speed. Consequently, the balance of therapeutic network coding benefits against network coding overheads must be worked out to sue for getting superior utilization in SDN-based computer networks. [2]

Analyzing the integration of network coding into Software-Defined Networking (SDN) topologies to address challenges related to packet loss, congestion, and fault tolerance is the major concern of this paper. Thus, it proposes a quantitative assessment of the impact of network coding on the performance indicators in SDN in terms of throughput, latency, packet loss rate, and fault tolerance. Furthermore, we examine the latency and the overhead of encoding and decoding to capture the cost and benefit of each. Although, network coding overcomes reliability and throughput challenges, it has an added disadvantage of incurring a processing delay from the encoding/decoding process.[2] [7]

In this paper, we evaluate the trade- off of implementing network coding, whereby we discuss the conditions under which this technique affords the maximum gains from inter-cluster communication while at the same time minimizing on the time delay. In light of these results, we gain important information regarding the real-world deployment of NC in SDN contexts. Thus, the goal of this study is to create foundation for building the dependable, efficient and flexible network system through integration of SDN and network coding.

In the subsequent sections, we provide further information regarding the background of SDN, its architectural components all together with its base principles, analyze challenges This research is driven by and provide the theoretical framework of network coding. Following that, we explain the details of our experimental configurations, procedures, and outcomes, along with a discussion of trade-offs and its application to practical scenarios. This work aligns with the research agenda of improving SDN features and presents data and analyses that can help network practitioners implement network coding as a method to improve network performance.

## II. Background and Related Work

Technological breakthroughs in networking have brought new ideas to meet these emerging needs, as well as, recognizing that networks are becoming more complex. Of these concepts, two have emerged as potential key enabling technologies namely, Software Defined Networking (SDN) and network coding. This section offers the reader with a brief understanding of the SDN architecture and how network coding works by giving an operating example.[8][9]

A. Software-Defined Networking (SDN) Architecture

Software Defined Networking (SDN) is a model that decouples the control plane from the forwarding plane where the control functions are centralized and can be programmed. Figure 1 explains the SDN architecture. In contrast to traditional networks, where control and data functions are tightly integrated within each device, SDN divides these responsibilities into three distinct layers:

1. Control Plane: In contrast, the control plane is utilized to plan over the necessary strategic choices from where the traffic is going to pass via the network. Located within an SDN controller, this layer computes routing algorithms, traffic policies, and its resource allocation strategies. Because the controller always has a global perspective of the network, he is able to fine tune the performance, ensure that policies are well implemented across the network, and easily deal with variations in traffic flow.

2. Data Plane: Data plane is composed of device such as switch and routers that blindly forwards packets based on the instructions from control plane. Such devices work in a way of just forwarding instructions with no component of decision making involved. This separation makes the device design much easier and also results in less operational complications.

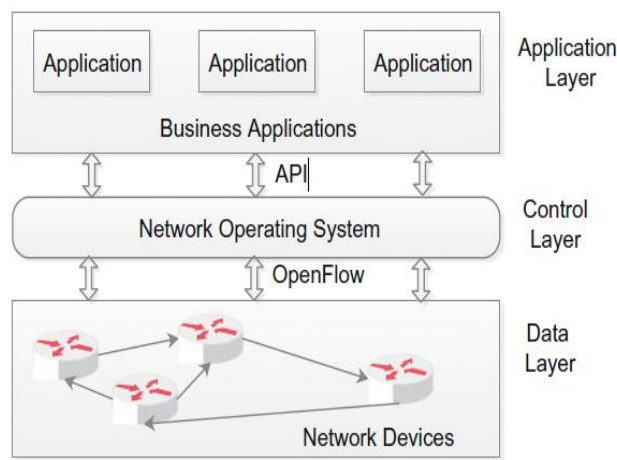

Fig. 1. SDN architecture

3. Application Programming Interfaces (APIs): Some APIs are used to enable interaction between control and data planes. SDN southbound API, such as OpenFlow integrate the controller to configure and manage forwarding devices, and the SDN northbound API, which allows applications to interface with the controller to describe network's behavior patterns. These APIs further improve modularity and scalability and make SDN highly programable.

However, problems like packet loss, congestion of the network, and limited tolerance to failure are areas that SDN will encounter in the near future in changed or failure environments.

B. Network coding

Network coding is a networking approach that would potentially enhance efficiency and data reliability by permitting middle-hoppers to encode transmitted packets. In contrast to conventional information networking, in which packets are disjointed entities, packets in network-coded packets are summed mathematically.[9]

1. Encoding and Decoding: At the source, data packets are sent and at intermediate nodes the details are together encapsulated in to the encoded packets through linear algebraic operations. The called recipients can then use the now coded packets therefore using their knowledge of the code scheme and uncoded the original data. This makes data communication more efficient and reduces the need of retransmission in the chance that packets have been lost.

2. Advantages in Multicast Scenarios: This is true because network coding is especially beneficial of scenarios where the same information is transmitted to several receivers; multicast. A further reduction in both distance and number of transmissions is achieved with the help of network coding that enables intermediate nodes to encode and forward packets. Through this optimization, there is enhancement of the bandwidth along with the deletion of latency.

3. Resilience to Packet Loss: In lossy environments for instance in wireless networks, the use of network coding improves on the reliability of transmitted data since the packets contain additional information about the transmitted data. Although some of the packets may be lost, the receiver is capable of reconstructing the information conveyed in the encoded packets thereby enhancing fault tolerance and through put.

The integration of network coding with software-defined networking results in the ability to control coding schemes from a central point that are capable of responding to the current conditions in the network and demand for traffic. This can help in overcoming the inherent intrinsic issues of SDN, and give a scalable, efficient and reliable network design.[9]

Figure 2 shows how the concept of network coding is implemented. Normally, if node A has information contents "a" and "b" to be sent to nodes B2 and C2 respectively, then the requirement of links (channels) between T1 and T2 are two as shown in figure 1 (left). In the network coding (NC) scheme, node T1 encodes the message it received as, for example, XOR of both "a" and "b" and is then able to forward this simultaneously to B2 and C2 through node G. As a result, B2 receives "a" and "a b" whereas C2 receives "b" and "a b." This means that both "a" and "b" can be gotten by solving a set of linear equations. Therefore, the bandwidth consumed through nodes T1 and T2 is cut by half.[9]

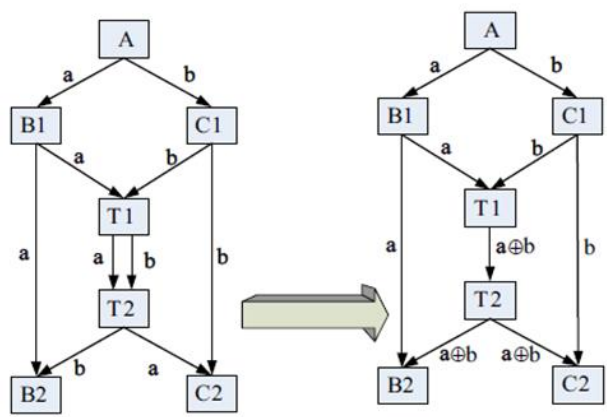

Fig. 2. The Network coding concept

For the related work, four articles where investigated. The first two papers related to SDN performance metrics. The others investigate previous work on applying network coding to networking problems.

The first paper compares distributed controller architectures in Software Defined Networking (SDN). The paper defines metrics for evaluating the performance of networks and categories different network structures according to graph properties. A first-cut performance prediction model is proposed in this paper using simple linear regression and future work will focus on the use of more sophisticated machine-learning-based predictive models. Nevertheless, there is a lack of research on distributed controller architectures across the different network topologies, inadequate modeling of interactions and side effects in relation to controller's placement and even inadequate predictive models which take into consideration the nonlinear nature of the relationship. Furthermore, there is no cross check of results on large scale test beds.[10]

The second paper evaluates the Quality of Service (QoS) parameters in Software Defined Networking, the relative experiment uses the Ryu controller with the Mininet emulator environment. It is best measured in bandwidth and round-trip time or RTT for short. There are other related network topologies that future research will focus on, but no concrete investigations have been made to different SDN controllers, other than Ryu. It also emphasizes the need for a comparative study of various load balancing approaches.[11]

The third provides a background on code-based routing for wireless networks so as to enhance performance and energy. The proposed method solves the problems in the current routing protocols to guarantee that network coding can be sustained even in cases where some nodes fail. There is little investigation of the coding options within the scope of routing metrics. Further, the categories of paths which should be selected and the ways for their selection needs be further developed, and also the effects of node failures on the routing. There is relatively poor coverage of the assessment of other coding schemes.[12]

The last article conclude that the use of network coding enhances the throughput as well as the performance of the wireless networks. It contains feature disadvantages like the delay in transmitting data and difficulty in coding. We also suppose transforming random network coding as sparse network coding lowers the computational complexity due to fewer zero coefficients. Unequal error protection protects crucial data most of the time by giving them precedence over other data. However, a major limitation with the deployment of network coding is the fact that conventional devices have severely constrained processing capabilities. Further, high decoding delays characteristic of the receiver end impact comprehensibility's of applications in actual applications.[13]

## III. METHODOLOGY

To evaluate the impact of network coding on SDN, this study focuses on key performance metrics: throughput, latency, packet loss, fault tolerance and load balancing. The methodology involves defining each metric and describing the process for measuring them mathematically under two scenarios: one with network coding and another one without network coding.

Throughput: This is the rate, expressed in Mbps (megabits per second) at which data are transferred in a given time period. It is an indication the communication channel efficiency.

Latency: The disruption of real time which corresponds to milliseconds during data transmission. Less amount of time means faster communication hence the meaning of low latency.

Packet Loss: The fraction of packets that are lost and do not actually reach their intended recipient. This is measure of the reliability of the system.

Fault Tolerance: The capacity of the system to perform well even if there are failures or errors in the system normally measured by the recovery rates of the system or system uptime.

Load Balancing: Proven by the fact that in the network, the traffic load is distributed evenly in every path which is solved by the variability of traffic load in channel or nodal.

table I shows the metrics and their equations with explanations and symbols descriptions for the two scenarios: with network coding and without network coding.

## IV. RESULTS AND DISCUSSION

In this section, the results highlighted to show how network coding can be integrated to offer benefits in terms of throughput, latency, packet loss, fault tolerance and load balancing in Software Defined Network (SDN). This evaluation is based on experiments made in a controlled SDN environment where different performance measurements have been made with and without network coding. Table II shows the results.

The base case saw throughput improve from 700 pps to 1,000 pps indicating an improvement of 42.8 percent. During high traffic load, the throughput improvement of network coding was showed almost twice as compared to no coding conditions thus showing that it is a useful method to solve congestion problems. These results demonstrate how network coding can optimize the utilization of network resources; most often in lossy or high load conditions.

In the base case the latency rose from 0.02 seconds to 0.025 seconds that points to a maximum of 25% increase. In high traffic and high loss conditions, the reduction of retransmission more than offsets the delay due to coding hence enhancement of latency in very high lossy environment. The above variation between the computational overhead of network coding and the consequent decrease of retransmission delay shows that the effects of network coding latency are tolerable and in fact, or more acceptable for high packet loss scenario networks.

TABLE I
EQUATION FOR THE METRICS EVALUATION

| Metric | Without coding | With coding | Explanation | Symbols description |
|---|---|---|---|---|
| Throughput (T) | $T_{uncoded} = \lambda * (1 - P_{loss})$ | $T_{coded} = T_{uncoded} * (1/(1 - P_{loss}))$ | Network coding increases throughput by recovering lost packets through redundancy | $\lambda$: Traffic load (packet/sec) $P_{loss}$: packet loss probability |
| Latency (L) | $L_{uncoded} = L_{request} + L_{prossing} + L_{response}$ | $L_{coded} = L_{uncoded} + L_{coding} + L_{reduced}$ | Network coding adds coding/decoding delay but reduces retransmission delays | $L_{coding}$: Dely introduced by encoding/decoding $L_{reduced}$: reduced retransmission delay |
| Packet Loss ($P_{loss}$) | $P_{loss,uncoded} = P_{loss}$ | $P_{loss,coded} = (P_{loss})^K$ | Coding reduces effective packet loss by introducing redundancy (K coded packets) | K: Redundancy factor (number of coded packets) |
| Fault Tolerance (R) | $R_{uncoded} = 1 - P_{failure}$ | $R_{coded} = 1 - (1- R)^K$ | Coding increase's fault tolerance by leveraging path diversity and ensuring recoverability from failures | $P_{failure}$: probability of path failure R: path reliability without coding |
| Load imbalance ($\Delta_{Imbalance}$) | $\Delta_{Imbalance,uncoded} = T_{max,uncoded} - T_{min,uncoded}$ | $\Delta_{Imbalance,coded} = T_{max,coded} - T_{min,coded}$ | Coding distributes traffic more evenly across multiple paths, reducing congestion on any single path | $T_{max}$: maximum load on a single path $T_{min}$: minimum load on a single path |
| Load Distribution ($T_{max}$) | $T_{max,uncoded} = \lambda * load$ fraction of heaviest path | $T_{max,coded} = \lambda/n$ | Traffic is evenly distributed across n paths with coding compered to uneven load in uncoded networks | N: number of paths in the network |

The base case analysis to show the reductions found the overall packet losses dropped from 30 % to 9 % or a reduction of 70 %. In high-loss scenarios packet loss was decreased from 60% to 36%. These results show a high reliability even in cases when messages delivered through a network coded environment have the potential to become corrupted.

TABLE II
THE FOUR CASESE EVALUATED RESULTS

| Case | Metric | Without Coding | With Coding | Effect |
|------|--------|----------------|-------------|--------|
| 1. Base Case | Throughput (packets/sec) | 700 | 1000 | Coding eliminates packet loss |
| | Latency (seconds) | 0.02 | 0.025 | Small delay from coding |
| | Packet Loss (%) | 30 | 9 | Loss reduced due to redundancy |
| | Fault Tolerance (%) | 70 | 91 | Increase reliability via coding |
| | Load Imbalance (%) | 60 | 20 | Balanced traffic from coding |
| 2. High Traffic ($\lambda$=2000) | Throughput (packets/sec) | 400 | 850 | Coding reduce congestion |
| | Latency (seconds) | 0.05 | 0.06 | Coding delay outweighs reduced retries |
| | Packet Loss (%) | 60 | 36 | Partial recovery from redundancy |
| | Fault Tolerance (%) | 60 | 84 | Coding mitigates failures |
| | Load Imbalance (%) | 80 | 40 | Better load balancing |
| 3. Long Distance (d=100km) | Throughput (packets/sec) | 650 | 950 | Coding improves flow despite delays |
| | Latency (seconds) | 0.04 | 0.05 | Increased delay from coding |
| | Packet Loss (%) | 35 | 15 | Loss mitigated through redundancy |
| | Fault Tolerance (%) | 68 | 88 | Increased robustness |
| | Load Imbalance (%) | 65 | 25 | Traffic evenly distributed |
| 4. High Path Failure ($P_{failure}$=0.5) | Throughput (packets/sec) | 600 | 850 | Coding compensates for failed paths |
| | Latency (seconds) | 0.03 | 0.035 | Slight delay increases from coding |
| | Packet Loss (%) | 50 | 25 | Improved reliability |
| | Fault Tolerance (%) | 50 | 75 | Recovery using multiple paths |
| | Load Imbalance (%) | 70 | 30 | Better traffic distribution |

Hence, fault tolerance increased from 50 percent in situations where there is no coding to 75 percent when coding was applied with regard to high-path failure. This improvement also highlights that the network coding has the capability of high reliability in path failure prone networks and especially, for dynamic or large scale SDN scenarios.

In the base case, therefore, load imbalance was reduced from 60% to 20% which also reduced the congestion of oversubscribed links. These results provide evidence of network coding in conditions of enhancing joint fairness and efficiency in traffic distribution with special reference to multi-path systems.

## IV. CONCLUSION AND FUTURE WORK

This is particularly true in wireless SDN and lossy data centers where the performance of network coding is very effective due to inherent packet losses. Despite the additional processing delay incurred by coding, the reduced chance of retransmission more than offsets the case of employing coding in real-time applications with other optimized coding algorithms. Further, the resulting better fault tolerance capability of network coding makes the approach suitable for dynamic networks where link or path failures are often observed. The implementation of network coding can also improve other aspects of SDN traffic engineering since load balancing and congestion are critical to any efficient network. These considerations illustrate the nonlinear signaling of network coding as an effective means for enhancing the SDN capability in multiple configurations.

Although the network coding approach offers superior performance on various evaluation criteria, the augmented computational load could negatively affect devices with limited communication capabilities in the future. Larger scale encoding and decoding of blocks will have overhead effect; future research should aim at improving encoding and decoding algorithms to minimize these effects. Besides, for the integration of network coding and distributed SDN controllers, many areas have not been investigated. Finer analysing the effect of network coding on against size of the plane of control and latency it would prove as a beneficial line of research. The further studies should also take into account the ability of coding schemes to adapt redundancy depending on the current conditions in the network, for instance traffic intensity or link failures. Also, trade-offs of energy utilization of introducing network coding in SDN for implementation for battery-operated or energy-sensitive devices are important.

## REFERENCES

[1] Kreutz, Diego, et al. "Software-defined networking: A comprehensive survey." Proceedings of the IEEE 103.1 (2014)

[2] Di, Jian, and Jingtao Dong. "A Network Coding architecture base on OpenFlow network." 2016 4th International Conference on Mechanical Materials and Manufacturing Engineering.

[3] Ali, Q. I. (2018). GVANET project: an efficient deployment of a self-powered, reliable and secured VANET infrastructure. IET Wireless Sensor Systems, 8(6), 313-322. DOI: 10.1049/iet-wss.2017.0189

[4] Ali, Q. I. (2012). Design and implementation of an embedded intrusion detection system for wireless applications. IET Information Security, 6(3), 171-182. DOI: 10.1049/iet-ifs.2011.0152

[5] Qaddoori, S. L., & Ali, Q. I. (2023). Efficient security model for IIoT system based on machine learning principles. Al-Rafidain Engineering Journal (AREJ), 28(1), 329-340.

[6] Ali, Q. I. (2009). Performance Evaluation of WLAN Internet Sharing Using DCF & PCF Modes. International Arab Journal of e-Technologies (IAJET), 1(1), 38-45.

[7] Lazim Qaddoori, S., Ali, Q.I.: An embedded and intelligent anomaly power consumption detection system based on smart metering. IET

Wirel. Sens. Syst. 13(2), 75–90 (2023).
https://doi.org/10.1049/wss2.12054

[8] Ali, Q. I. (2016). Green communication infrastructure for vehicular ad hoc network (VANET). Journal of Electrical Engineering, 16(2), 10-10.

[9] Gu, Rentao, et al. "Efficient software-defined passive optical network with network coding." Photonic Network Communications 31 (2016)

[10] Gray, Nicholas, Katharina Dietz, and Tobias Hossfeld. "Simulative evaluation of KPIs in SDN for topology classification and performance prediction models." 2020 16th International Conference on Network and Service Management (CNSM). IEEE, 2020.

[11] Babbar, Himanshi, and Shalli Rani. "Performance evaluation of qos metrics in software defined networking using ryu controller." IOP conference series: materials science and engineering. Vol. 1022. No. 1. IOP Publishing, 2021.

[12] Rajkumar, M., R. Radhika, and J. Karthika. "Code-Based Routing Using Distributed Approach for Wireless Networks." 2020 3rd International Conference on Intelligent Sustainable Systems (ICISS). IEEE, 2020.

[13] Zhu, Fumin, et al. "Practical network coding technologies and softwarization in wireless networks." IEEE Internet of Things Journal 8.7 (2021): 5211-5218.

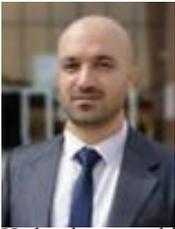

**Amer T. Ali**, He completed his Bachelor of Science in Computer Engineering from Mosul University, Mosul Iraq in the year 2006. The same year he completed a second degree, MSc degree in Computer Engineering at the same university. At the present time, he is working toward a Ph.D. degree in Computer Engineering in the College of Engineering, Mosul University. He has been working at Nineveh University in Iraq even from 2012 till now. His area of specialization is computer architecture and channel coding for networking.